\documentclass{ws-mpla}
\usepackage[super]{cite}
\usepackage{graphicx}
\usepackage{amsmath,amssymb}
\usepackage{slashed}
\begin{document}

\markboth{Dmitri Vassilevich}
{Casimir repulsion between Chern-Simons surfaces}

\title{ON THE (IM)POSSIBILITY OF CASIMIR REPULSION BETWEEN CHERN-SIMONS SURFACES}

\author{\footnotesize DMITRI VASSILEVICH}

\address{CMCC - Universidade Federal do ABC, Avenida dos Estados 5001, CEP 09210-580, \\ Santo Andr\'e, SP, Brazil \\
Department of Physics, Tomsk State University, 634050 Tomsk, Russia
\\
dvassil@gmail.com}

\maketitle

\begin{abstract}
We critically review the possibilities of a repulsive Casimir interaction between materials which carry Chern-Simons surface actions for the electromagnetic field. 
\end{abstract}

\section{Introduction: the Chern-Simons action}
The Chern-Simons (CS) action appeared 45 years ago\cite{Chern:1974ft} as a boundary term preventing a simple combinatorial interpretation of the Pontryagin number. Since that time, this action has found a lot of applications in various areas of physics. For an electromagnetic potential $A_j$ in a $(2+1)$-dimensional Minkowski space the CS action reads
\begin{equation}
S_{\rm CS}=\frac {ke^2}{4\pi} \int d^3x\, \varepsilon^{ijl}A_i\partial_jA_l\,,\label{CS}
\end{equation}
where $\varepsilon^{ijl}$ is the Levi-Civita tensor, $e$ is the elementary charge. The dimensionless constant $k$ is called the CS level.

By varying $S_{\rm CS}$ with respect to $A_i$, one can compute the CS current $j^i$. It is easy to see, that this current is always perpendicular to the electric field. Thus (\ref{CS}) describes a Hall type conductivity, which hints to numerous applications in the condensed matter physics.

The CS action with $|k|=\tfrac 12$ is the parity anomaly of a Dirac fermion in $(2+1)$ dimensions.\cite{Niemi:1983rq,Redlich:1983dv} In other words, the parity odd part of the quantum effective action of a single Dirac fermion in $(2+1)$ dimensions is the CS action with $k=\pm\tfrac 12$. 

Besides, the CS action describes a Topological Field Theory and a Conformal Field Theory. The study of these aspects of CS theory has led to many interesting results in topology of 3-manifolds, invariants of knots and other areas of mathematics and mathematical physics.

Therefore, the influence of CS action on the Casimir effect is a very natural problem to address. This study started long ago out of curiosity. The situation changed with the discovery of Dirac materials where elementary excitations satisfy the quasirelativistic Dirac equation and thus bring quantum anomalies into the game. Besides purely theoretical interest, CS terms on the boundary opened a possibility to obtain a repulsive Casimir force -- a phenomenon having many potential technological applications. 

This micro review is organized in a chronological order. We shall start with the very first papers and continue with more and more recent and sophisticated models that describe Dirac materials increasingly well. While discussing these models we shall be mostly interested in the possibility of a repulsive Casimir interaction. 

\section{Casimir interaction of Chern-Simons surfaces}
One possibility to study is the Casimir interaction an electromagnetic action given by a sum of 
Maxwell and CS for the photons in the bulk of a $(2+1)$ dimensional space.\cite{Milton:1990yj} However, it is more natural from the physical point of view to keep the Maxwell action in a $(3+1)$ dimensional bulk and put the CS term on a $(2+1)$ dimensional boundary. Such a system was initially considered\cite{Elizalde:1998ha} with the purpose to clarify some aspects of the heat kernel expansion with oblique boundary conditions. The Casimir interaction was computed in Ref.\ \citen{Bordag:1999ux}. In the set-up of both papers\cite{Elizalde:1998ha,Bordag:1999ux}, the CS interaction modified rigid dual superconductor boundary conditions. Although this set-up is hard to implement in a realistic physical situation, the paper\cite{Bordag:1999ux} contained an important message: under certain conditions the Casimir force between two surfaces that carry CS interactions may become repulsive.

A different approach was adopted in Ref.\ \citen{Markov:2006js}. The CS action was placed on a semitransparent surface, so that the CS term defined the matching conditions for electromagnetic field rather than modified the rigid (impenetrable) boundary conditions. Again, a repulsive Casimir force between surfaces caring CS actions was found that in certain range of the coupling. Some years later it was realized that this model provides a (rather simplified) description of topological insulators (see, e.g., Ref.\ \citen{Tkachov2013}), which triggered extensive studies in this direction.\cite{Grushin:2011,Grushin:2011b,RodriguezLopez:2011dq,Nie:2013,Marachevsky:2017tdo,Fialkovsky:2018fpo,Marachevsky:2018axn}

Most of the papers mentioned in the previous paragraph considered the Casimir interaction of two bodies having a dielectric bulk and a Hall conductivity (CS term) on the surface. The CS term on the surface of topological materials is induced through quantum effects of specific states localized near the surface, as we sketched in the previous section. However, the same localized states also induce a longitudinal conductivity. To estimate the influence of this conductivity, let us consider the surfaces that carry a constant conductivity\cite{Fialkovsky:2018fpo}
\begin{equation}
\sigma_{ij}=2(\zeta \delta_{ij}+\eta \varepsilon_{ij}) \label{sigma}
\end{equation}
containing a longitudinal term proportional to the unit matrix $\delta_{ij}$ as well as a Hall term with a 2D Levi-Civita tensor $\varepsilon_{ij}$. The Casimir repulsion is only possible if the Hall conductivities of interacting surfaces have the same sign.\footnote{The claims of Casimir repulsion between surfaces with Hall conductivities of opposite signs have been attributed\cite{Fialkovsky:2018fpo} to the use of an incorrect version of the Lifshitz formula.}

\begin{figure}[ph]
\centerline{\includegraphics[width=3.0in]{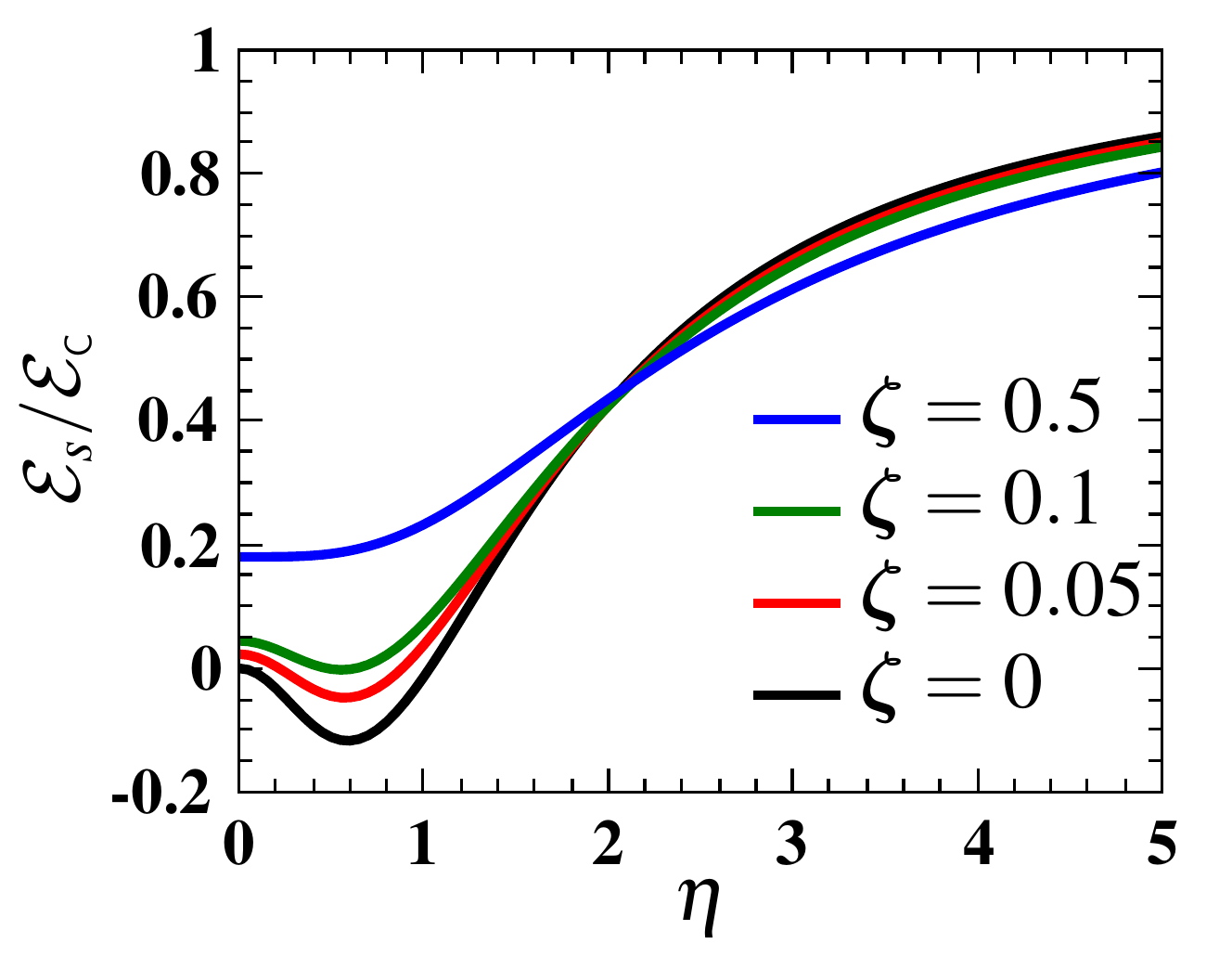}}
\vspace*{8pt}
\caption{(Color online) The energy per unit area $\mathcal{E}_s$ of Casimir interaction of two surfaces with equal conductivities (\ref{sigma}) normalized to the Casimir energy $\mathcal{E}_C$ of two ideal conductors. This figure is taken from Ref.\ \citen{Fialkovsky:2018fpo}.  \protect\label{fig1}}
\end{figure}

To get an idea of the strength of the effect we depicted, Fig.\ \ref{fig1}, the dependence of the Casimir energy density for two parallel plates carrying the same constant conductivity (\ref{sigma}) on the Hall parameter $\eta$ for several values of the longitudinal conductivity. The energy density is normalized to the corresponding quantity for two ideal conductors separated by the distance $a$:
\begin{equation}
\mathcal{E}_C=-\frac{\pi^2}{720a^3}\,. \label{EC}
\end{equation}

The curve for $\zeta=0$ reproduces the results of the paper.\cite{Markov:2006js} We see, that for moderate values of the conductivities a Casimir repulsion is possible. The strongest repulsive force is obtained for $\eta\simeq 0.5$, which is the values used by Marachevsky.\cite{Marachevsky:2018axn} Can this value of the coupling be achieved in topological materials? The Hall conductivity constant $\eta$ is related to the CS level through the formula
\begin{equation}
\eta = \frac{ke^2}{4\pi} \,.\label{etak}
\end{equation}
One fermion mode in $(2+1)$ generates $|k|=\tfrac 12$, as we mentioned in the previous section. To achieve the value $\eta\simeq 0.5$ one needs more than one hundred of surface modes, which is just not realistic. For a small number of surface modes, the parameter $\eta$ is also small. Even more important, the longitudinal dc conductivity $\zeta$ of $(2+1)$ dimensional fermions has the same order of magnitude as the Hall conductivity $\eta$. For small conductivities, the Casimir force depends quadratically on $\eta$, see Fig.\ \ref{fig1}. On the contrary, the longitudinal conductivity leads to Casimir attraction and the dependence on $\zeta$ is linear (as one can see  from the explicit formulas\cite{Khusnutdinov:2015:cefasocp} or guess from Fig.\ \ref{fig1}). The longitudinal conductivity always takes over the Hall part leading to an overall attractive Casimir force.

\section{Topological insulators: more sophisticated models}
One rather obvious improvement of the model described in the previous section consists in replacing constant conductivities by the frequency and momentum dependent conductivities defined through the polarization tensor of $(2+1)$ dimensional Dirac particles. The polarization tensor approach to the Casimir interaction of graphene was suggested in Refs.\ \citen{Bordag:2009fz,Fialkovsky:2011pu}. The analysis\cite{Klimchitskaya:2014axa} demonstrated that the polarization tensor approach has the best agreement with Casimir experiment.\cite{Banishev:2013} The principal difference between the situation considered here and the one in graphene is that in graphene the parity odd part of the polarization tensor is cancelled between contributions of various generations of quasiparticles, while here this part is of the main interest.

In the momentum representation the parity odd part of one-loop effective action of a single Dirac fermion in $(2+1)$ dimensions with mass $m$ reads\cite{Appelquist:1986fd}
\begin{equation}
S_{\rm odd}=\frac {e^2}{8\pi}\int \frac{d^3p}{(2\pi)^3} \varepsilon^{jkl} A_j(-p)(i p_k)A_l(p) \left[ \frac{2m\, \mathrm{arctanh}(\tilde p/2m)}{\tilde p} -\mathbb{A} \right],\label{Sodd}
\end{equation}
where $\tilde p=\sqrt{p_0^2-v_F^2(p_1^2+p_2^2)}$, and $v_F$ is the Fermi velocity that varies between $10^{-3}$ and $10^{-2}$, depending on the material. The parameter $\mathbb{A}$ describes the contribution of parity anomaly. If $\mathbb{A}=1$ the parity anomaly is included, while if $\mathbb{A}=0$, it is not. Note, that the overall sign in front of $S_{\rm odd}$ may be inverted by choosing a different inequivalent representation for the $\gamma$-matrices. Thus in the $m\to 0$ limit one can obtain the CS action with the level $k\pm\tfrac 12$. 

\begin{figure}[ph]
\centerline{\includegraphics[width=3.0in]{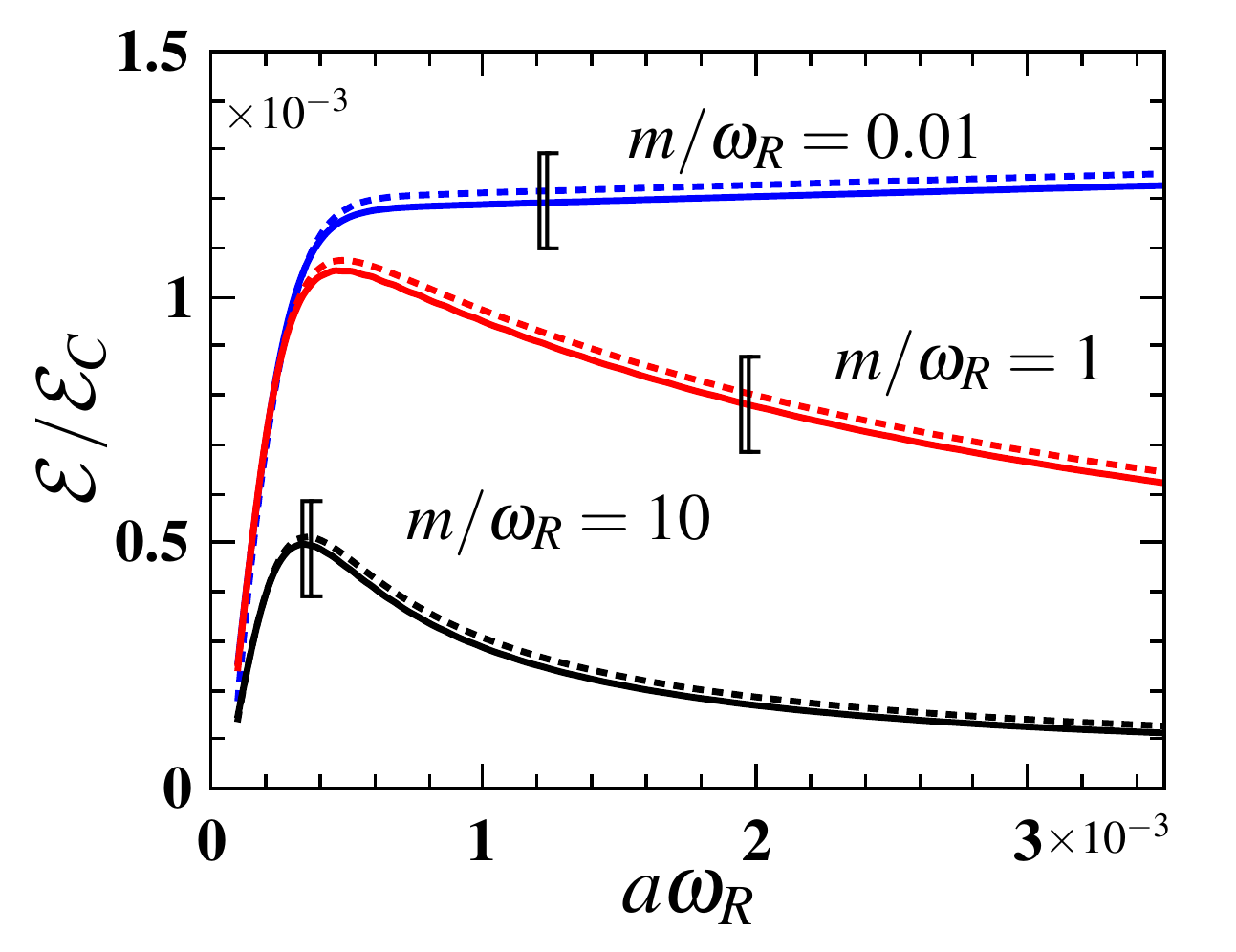}}
\vspace*{8pt}
\caption{(Color online) The energy per unit area $\mathcal{E}$ of Casimir interaction of two dielectric bodies that carry surface modes of equal masses and with identical (solid lines) and opposite (dashes lines) signs in front of the parity odd actions normalized to the Casimir energy $\mathcal{E}_C$ of two ideal conductors. Dimensionless distance $a\omega_R$ and mass $m/\omega_R$ are used (with $\omega_R$ being the resonant frequency of the dielectric bulk). This figure is taken from Ref.\ \citen{Fialkovsky:2018fpo}.  \protect\label{fig2}}
\end{figure}

The Casimir interaction of topological insulators with surface conductivities modelled by contributions of massive particles in $(2+1)$ dimensions was computed in the papers,\cite{Chen:2011,Chen:2012ds} where, however, the parity anomaly was neglected.

There are two basic arguments in favour of the parity anomaly. First, this anomaly is necessary to support the invariance of fermion determinant with respect to large gauge transformations. Second, the presence of parity anomaly ensures decoupling of massive modes in the parity odd effective action. Indeed, the action (\ref{Sodd}) vanishes in the limit $m\to\infty$ if and only if $\mathbb{A}=1$. While validity of the first argument in the context of condensed matter physics is questioned by some people, the second argument is definitely relevant: very massive particles should not contribute to the conductivity. Neglecting the anomaly, the authors\cite{Chen:2011,Chen:2012ds} came to a conclusion that the Casimir repulsion appears if the mass is larger than some threshold, which looks strange from the physical point view. In the paper\cite{Fialkovsky:2018fpo} the Casimir interaction was computed in the same model as in the works\cite{Chen:2011,Chen:2012ds} but including the parity anomaly. Some results are presented on Fig.\ \ref{fig2}. The force always remains attractive. Moreover, the sign of the parity odd action does not influence the Casimir interaction to any significant degree. Qualitatively, this results confirms the conclusions drawn from the model with constant conductivities.

Another relevant question is how well the electronic properties of topological insulators are described by $(2+1)$ dimensional fermions localized on the surface? At a more fundamental level, the quasiparticles in Dirac materials obey the Dirac equation in the bulk and some boundary conditions on the surface. To be precise, let us take the Dirac operator in the form
\begin{equation}
\slashed{D}=i\tilde\gamma^\mu \bigl( \partial_\mu +i e A_\mu) +i m_5\gamma^5 +m ,\label{Dir}
\end{equation}
where $\tilde\gamma^\mu=\eta^\mu_\nu \gamma^\nu$ with $\eta \equiv \mathrm{diag}(1,v_F,v_F,v_F)$.
 The simplest set of local boundary conditions that ensure vanishing normal current on the boundary $\partial\mathcal{M}$ is the bag conditions\cite{Chodos:1974je}
\begin{equation}
(1\pm i\gamma^n)\psi\vert_{\partial\mathcal{M}} =0 ,\label{bagbc}
\end{equation}
where $\gamma^n =n^\mu \gamma_\nu$, and $n^\mu$ is an inward pointing unit normal to the boundary. Note, that both signs in front of $i\gamma^n$ are allowed.

Our purpose here is to compare the effective action for external electromagnetic field of quantized fermions described by (\ref{Dir}) and (\ref{bagbc}) to simplified models with purely $(2+1)$ dimensional modes.

If $m_5=0$ and the sign in (\ref{bagbc}) conspires with the sign of $m$, there is a gapless surface mode of the Dirac operator that is localized near the boundary with the characteristic localization length  $|m|^{-1}v_F$. The presence of this mode does not depend on the shape of the boundary. Thus, this mode is \emph{topologically protected}. A non-zero $m_5$ gives to this mode a mass gap.

In the massless case, $m=m_5=0$, on Euclidean manifolds the parity odd part of the effective action can be computed with the heat kernel methods exactly for any shape of the boundary yielding\cite{Kurkov:2017cdz} the CS action on the boundary of the manifold with $k=\pm 1/4$. Since there are no masses, this case corresponds to a Dirac semimetal rather than to a topological insulator.

For non-zero values of $m$ and/or $m_5$ the heat kernel methods do not give exact expressions for the parity odd effective action. Thus, one should rather use the standard Feynman diagrams for which one needs explicit expressions for the propagators. As a consequence one has to restrict computations to some simple geometry, let us say a half-space. One important technical message of this computation\cite{Fialkovsky:2019rum} is that the Pauli-Villars scheme fails to regularize the theory unless one gives axial masses to the regulator fields. The effective action is not localized on the boundary, though one can define a boundary contribution by integrating relevant parts over normal coordinates near the boundary. In the parity odd sector one obtains an action similar to (\ref{Sodd}), but with a different form-factor depending on both $m$ and $m_5$. Exact expressions may be found in.\cite{Fialkovsky:2019rum} Roughly speaking, the near boundary Hall type conductivity is always smaller than the corresponding conductivity of a $(2+1)$ dimensional Dirac fermion.  

One can generalize the boundary conditions by inserting a chiral factor,
\begin{equation}
(1\pm ie^{i\theta\gamma^5}\gamma^n)\psi\vert_{\partial\mathcal{M}} =0 ,\label{chbag}
\end{equation}
depending on a constant chiral angle $\theta$. These conditions are called the chiral bag boundary conditions.\cite{Rho:1983bh} The CS part of the effective action was evaluated\cite{MateosGuilarte:2019eem} by comparing to the fermion number of domain walls. No enhancement of the Hall conductivity as compared to the $\theta=0$ case was observed.

Thus, more realistic models of the Dirac materials do not give an enhancement of the surface Hall conductivity sufficient to produce a Casimir repulsion between such materials.
 
\section{Some conclusions}
We see, that after the period of initial enthusiasm a more sober consideration based on realistic models of Dirac materials uncovered two basic difficulties in obtaining a repulsive Casimir force between these materials. First, the surface Hall conductivity appears too small. Second, the longitudinal surface conductivity has a larger effect causing an attractive force.   

We should stress that these conclusion does not refer to the Casimir interaction when the Hall conductivity is caused by external magnetic field\cite{Tse:2011} or when the CS term is presented in the bulk rather than on the boundary (see\cite{Fukushima:2019sjn} and references therein). At any rate, one may expect further interesting developments at the interface between the physics of new materials and the Casimir effect.\cite{Woods:2015pla}

\section*{Acknowledgments}
The author is grateful to the organizers of 4th Symposium on the Casimir effect for making such a fruitful and stimulating meeting. He also thanks Michael Boradg, Ignat Fialkovsky, Nail Khusnutdinov and Maxim Kurkov for collaboration on the topics presented in this contribution.
This work was supported in parts by the S\~ao Paulo Research Foundation (FAPESP), projects 2016/03319-6 and 2017/50294-1 (SPRINT), by the grants 303807/2016-4 and 428951/2018-0 of CNPq, by the RFBR project 18-02-00149-a and by the Tomsk State University Competitiveness Improvement Program.

\bibliographystyle{ws-mpla}
\bibliography{graphene}

\end{document}